\documentclass[prb,twocolumn,showpacs,superscriptaddress,amsmath,amssymb]{revtex4}
\usepackage{graphicx}

\usepackage{dcolumn}
\usepackage{bm}

\begin{document}

\title{Dynamics of photoexcited states in one-dimensional dimerized Mott insulators}

\author{Nobuya Maeshima}
\email{maeshima@ims.ac.jp} 
\homepage{http://magellan.ims.ac.jp/~maeshima/}
\affiliation{Department of Chemistry, Tohoku University, Aramaki, Aoba-ku, Sendai 980-8578, Japan}
\affiliation{Institute for Molecular Science, Okazaki 444-8585, Japan}

\author{Kenji Yonemitsu}
\affiliation{Institute for Molecular Science, Okazaki 444-8585, Japan}
\affiliation{Department of Functional Molecular Science, Graduate University for Advanced Studies, Okazaki 444-8585, Japan}

\date{\today}

\begin{abstract}
Dynamical properties of photoexcited states are theoretically studied in a one-dimensional Mott insulator dimerized by the spin-Peierls instability.  Numerical calculations combined with a perturbative analysis have revealed that the lowest photoexcited state without nearest-neighbor interaction corresponds to an interdimer charge transfer excitation that belongs to dispersive excitations.  This excited state destabilizes the dimerized phase, leading to a photoinduced inverse spin-Peierls transition.  We discuss the purely electronic origin of midgap states that are observed in a latest photoexcitation experiment of an organic spin-Peierls compound, K-TCNQ (potassium-tetracyanoquinodimethane).
\end{abstract}


\maketitle

\section{Introduction}

Photoirradiation is one of important methods to control physical properties of strongly correlated electron systems.~\cite{yu,PrMnO3,TTFCA,poly,MX,MMX,TTTA,EDO,sitaET,KTCNQ1,KTCNQ2}   Drastic changes of macroscopic phases triggered by the photoirradiation, often called ``photoinduced phase transitions (PIPTs)'',~\cite{nasu,yonemitsu1} are observed in a variety of materials where electrons are coupled with lattice systems.~\cite{TTFCA,poly,MMX,TTTA,EDO,sitaET,KTCNQ1}

Many PIPTs are understood in the framework of the adiabatic approximation.  The key quantities are adiabatic potentials of few electronic states as a function of a lattice configuration (represented by $Q$).  The adiabatic potential of the photoexcited electronic state has an energy minimum at $Q=Q^e_{\rm min}$ that is generally very different from the stable lattice configuration $Q^g_{\rm min}$ in the adiabatic potential of the ground state.   According to the Franck-Condon theorem, the lattice configuration of the photoexcited state is initially at $Q^g_{\rm min}$ ( {\it i.e.}, immediately after the photoexcitation from the ground state).  Then, the lattice configuration moves from $Q^g_{\rm min}$ to  $Q^e_{\rm min}$  along the adiabatic potential of the photoexcited state.  Such a photoinduced transformation of the lattice configuration takes place macroscopically in a PIPT for electron systems coupled with lattice systems.~\cite{nasu,hanamura,nagaosa1,koshino1}

The understanding of why the photoexcited state has a stable point different from that in the ground state is essential to reveal the fundamental physics of the PIPTs.  To address this problem theoretically, we should first clarify how the photoexcitation induces charge transfer (CT) processes, although it is generally difficult to carry out in strongly correlated electron systems.~\cite{huai,yonemitsu2,iwano,yonemitsu3}

As a typical example of strongly correlated electron systems coupled with lattice systems, we focus on a one-dimensional (1D) electron system with a dimerized lattice distortion, which is realized in an organic material, potassium-tetracyanoquinodimethane (K-TCNQ).  This material is a 1D dimerized Mott insulator below the spin-Peierls transition temperature, $T_{\rm sP}=395$K.~\cite{vegter,sakai,terauchi}  In 1991, Koshihara {\it et al.} have demonstrated that an irradiation of a pulsed laser weakens the lattice dimerization of K-TCNQ in the dimerized phase, which is regarded as a photoinduced inverse spin-Peierls transition.~\cite{KTCNQ1}
More recently, a high-resolution experiment has revealed several ultra-fast phenomena immediately after the photoirradiation, for example, the emergence of a midgap state in the reflectivity spectrum within the time resolution of 150fs and its decay within about 10 ps.~\cite{KTCNQ2}

In this paper, we clarify how the photoinduced CT excitation occurs in the 1D dimerized Mott insulator, by employing the 1D dimerized Hubbard model coupled with a uniform lattice distortion.  By the exact diagonalization method and a perturbative calculation, we demonstrate that an interdimer CT state, which is the lowest photoexcited state, destabilizes the dimerized phase.  The interdimer CT state is found to be superposition of two dispersive elementally excitations.  We investigate the origin of the midgap state observed experimentally in K-TCNQ.~\cite{KTCNQ2}  Candidates for the midgap state are shown to be intradimer and interdimer CT excitations from the photoexcited state (not from the ground state).  Relevance to K-TCNQ is also discussed.

\section{model}

\begin{figure}[hbt]
\begin{center}
\includegraphics[width=6.0cm,clip]{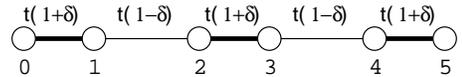}
\end{center}
\caption{Dimerized 1D system with alternating transfer integrals.  The thick lines correspond to the larger transfer integrals, and the thin lines to the smaller ones.}
\label{fig:dimer}
\end{figure}

To treat the dimerized phase of the 1D strongly correlated electron system coupled with the lattice system, we use a model whose Hamiltonian is given by
\begin{equation}
{\cal H} = {\cal H}_{\rm e} + \frac{N}{2}\beta \delta^2,
\end{equation}
where ${\cal H}_{\rm e}$ is for the electronic part and the second term gives the energy of the lattice system.  The elastic constant of the lattice is denoted by $\beta$, and $\delta$ $(0\le\delta\le1)$ is the lattice dimerization parameter.  The system size is denoted by $N$.  Taking the adiabatic approximation for the lattice system, we omit the kinetic energy term of the lattice.  The electronic part ${\cal H}_{\rm e}$ is the 1D dimerized Hubbard model defined by
\begin{eqnarray}
{\cal H}_{\rm e} =
 &-&\sum_{\sigma}\sum_{l=0}^{N-1}t[1+(-1)^l\delta]
( c^\dagger_{l+1,\sigma}c^{}_{l,\sigma} +  c^\dagger_{l,\sigma}c_{l+1,\sigma} )
\nonumber \\ &+& U\sum_l n_{l,\uparrow}n_{l,\downarrow}, \label{eq:ham}
\end{eqnarray}
where $c^{\dagger}_{l,\sigma}$ ($c_{l,\sigma}$) is the creation (annihilation) operator of an electron with spin $\sigma$ at site $l$, $n_{l,\sigma}=c^{\dagger}_{l,\sigma}c_{l,\sigma}$. 
The nearest-neighbor transfer integrals alternate as $t(1+\delta),t(1-\delta),\cdots$, as shown in Fig.~\ref{fig:dimer}.  The parameter $t$ gives the transfer integral of the uniform lattice, and is used as the unit of energy in this paper.  The filling of electrons is set to the half and the boundary condition is set to be periodic.  The Coulomb interaction $U/t$ of K-TCNQ has been estimated to be about $6-18$.~\cite{yakushi,meneghetti,gallagher}  Almost all results shown in our paper are obtained for $U/t=15$.

\section{Numerical method}

To search important photoexcited states, we calculate the optical conductivity spectrum of the ground state $|\psi_{0}\rangle$ given by~\cite{shastry}
\begin{equation}
\sigma(\omega)\equiv D\delta(\omega) +\sigma^{\rm reg}(\omega),
\label{eq:sigma0}
\end{equation}
where $D$ is its Drude weight defined by
\begin{equation}
D = - \frac{\pi}{N} \langle\psi_0|K|\psi_0\rangle - \frac{2\pi}{N}\sum_{n > 0}\frac{|\langle \psi_{ n}|J|\psi_0\rangle|^2}{E_{n}-E_0},
\end{equation}
with $K$ being the kinetic term of the Hamiltonian~(\ref{eq:ham}), $J$ the current operator defined by
\begin{equation}
J\equiv it\sum_{l,\sigma}[1+(-1)^l\delta]( c^\dagger_{l+1,\sigma}c_{l,\sigma} -  c^\dagger_{l,\sigma}c_{l+1,\sigma}),
\end{equation}
$E_0$ the ground state energy, $|\psi_n\rangle$ the $n$-th excited state, and $E_n$ the corresponding energy. The lattice constant is set to be unity in this work.  The regular component $\sigma^{\rm reg}(\omega)$ is defined by
\begin{equation}
\sigma^{\rm reg}(\omega)= -\frac{1}{N\omega}{\rm Im}\left[\langle\psi_{\rm 0}|J\frac{1}{\omega+i\epsilon+E_{\rm 0} -{\cal H}}J|\psi_{\rm 0}\rangle \right],
\label{eq:sigma-reg}
\end{equation}
where $\epsilon$ yields finite broadening and is set at 0.1$t$ in our calculations below.  We carry out the full diagonalization to obtain precise excited states for $\sigma(\omega)$ and other quantities.  Thus the system size $N$ is restricted to a relatively small value of $N=8$.

\section{numerical results}

\subsection{Optical conductivity}

\begin{figure}[hbt]
\begin{center}
\includegraphics[width=7.0cm,clip]{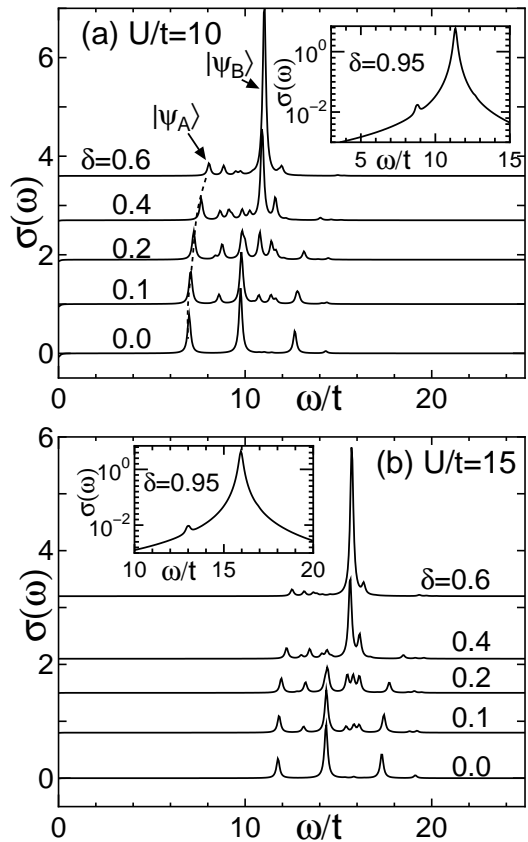}
\end{center}
\caption{Optical conductivity spectra in the 1D dimerized Hubbard model for $U/t=10$ and 15.  The insets show results of nearly decoupled systems ($\delta=0.95$).}
\label{fig:sw_g}
\end{figure}

Figure~\ref{fig:sw_g} shows the optical conductivity spectra $\sigma(\omega)$ of the model~(\ref{eq:ham}) for $U=$ 10 and 15.  In both cases, $\sigma(\omega)$ has similar characteristics mentioned below.  At $\delta=0$, there are a few states spread over $U-4t<\omega<U+4t$.  It is well known that they consist of holon-doublon pair excitations, forming the so-called holon-doublon continuum in the $N\to \infty$ limit.~\cite{fye,stephan,jeck}
With finite $\delta$ introduced, there appear several peaks with small spectral intensity, suggesting that new optically-allowed states are generated by breaking of the translation symmetry.  Accordingly, the photoexcited states at $\delta=0$ get weakened spectral intensity and some of them eventually disappear.
However, the lowest photoexcited state survives up to the decoupling limit (see the insets of Fig.~\ref{fig:sw_g}).  We focus on this state as a typical example of the holon-doublon states, and call this $|\psi_{\rm A}\rangle$ in the following.

As $\delta$ is further increased ($\delta>0.4$), a new optical state, which is called $|\psi_{\rm B}\rangle$ hereafter, appears as an exciton-like sharp peak slightly above $\omega=U$.  The appearance of this sharp peak is predicted in several preceding studies.~\cite{campbell,gallagher,lyo,gebhard1,gebhard2}
 This exciton-like state is found to belong to a different class from the holon-doublon continuum.  Thus we focus on the two states, $|\psi_{\rm A}\rangle$ and $|\psi_{\rm B}\rangle$, as typical photoexcited states in the dimerized system.  The nature of these two states is discussed in the following.

\subsection{Adiabatic potentials}

\begin{figure}[hbt]
\begin{center}
\includegraphics[width=7.0cm,clip]{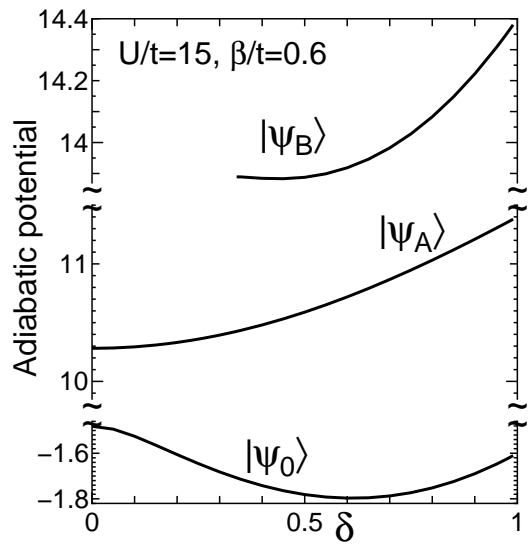}
\end{center}
\caption{Adiabatic potentials $E_i(\delta)$ of three states ($i=$0, A, and B) for the $N=8$ system. }
\label{fig:adptls}
\end{figure}

In this subsection, we show the adiabatic potentials of the relevant states, $|\psi_0\rangle$, $|\psi_{\rm A}\rangle$ and  $|\psi_{\rm B}\rangle$, to discuss their spin-Peierls stability.  The adiabatic potential $E_i({\delta})$ is given by
\begin{equation}
E_i(\delta)=E^{\rm e}_i(\delta)+\frac{N}{2}\beta\delta^2,
\end{equation}
where $E^{\rm e}_i(\delta)$ denotes the energy of the eigenstate $|\psi_i\rangle$ of ${\cal H}_{\rm e}$ ($i$=0, A, or B).  Figure~\ref{fig:adptls} shows the results for $U=15$.  The adiabatic potential of $|\psi_{\rm B}\rangle$ is plotted only in the region where $|\psi_{\rm B}\rangle$ has a pronounced spectral peak in $\sigma(\omega)$.   The elastic coupling $\beta/t$ is set to 0.6 here for the adiabatic potential of $|\psi_0\rangle$ to have a minimum at a considerably large value of the dimerization $(\delta\sim 0.6)$ to help us observe dimerization effects clearly.

We have found that $E_{\rm A}(\delta)$ has a minimum at $\delta=0$, which demonstrates that the dimerized state is destabilized by the optical excitation from $|\psi_0\rangle$ to $|\psi_{\rm A}\rangle$.   Hence, when the state $|\psi_{\rm A}\rangle$ is photoexcited, the system can move toward the uniform state ($\delta=0$) along the adiabatic potential $E_{\rm A}(\delta)$ unless the recombination of the holon and the doublon occurs; this is the realization of the inverse spin-Peierls transition.  By contrast, $E_{\rm B}(\delta)$ has a minimum around $\delta=0.4$.  This suggests that  $|\psi_{\rm B}\rangle$ does not annihilate the lattice dimerization although it can weaken it.

\section{decoupling limit}

In this section, we present a simple picture of the relevant photoexcited states $|\psi_{\rm A}\rangle$ and $|\psi_{\rm B}\rangle$ to understand their nature.  For this purpose we introduce a perturbation theory that is justified in the decoupling limit $\delta\to 1$.  Since both of $|\psi_{\rm A}\rangle$ and $|\psi_{\rm B}\rangle$ have finite spectrum intensity in the nearly decoupled system, it is reasonable to deal with both states from the decoupling limit, where the perturbative treatment of interdimer interactions is valid.

Several studies have presented different approximations for the 1D dimerized Hubbard model to discuss the characteristics of $\sigma(\omega)$, for example, the dominant spectral intensity around $\omega\sim U$.~\cite{gallagher,lyo,gebhard1,gebhard2}
However, these preceding theories are justified only in the $U/t\to\infty$ limit.  By contrast, our approximation can treat $U/t$ exactly, and is proved to be appropriate to construct a quite simple picture of the photoexcited states.

\subsection{Hamiltonian}

In the decoupling limit ($\delta\to 1$), the electronic part ${\cal H}_e$ can be categorized into the intradimer part ${\cal H}_0$ and the interdimer part ${\cal H}_1$ as follows:
\begin{equation}
{\cal H}_e= {\cal H}_0 + {\cal H}_1,
\end{equation}
with
\begin{eqnarray}
{\cal H}_0 &=& -t(1+\delta)\sum_{\sigma}\sum_{l\in \rm e}
[ c^\dagger_{l+1,\sigma}c_{l,\sigma} +  c^\dagger_{l,\sigma}c_{l+1,\sigma} \nonumber \\
 &&+  U (n_{l,\uparrow}n_{l,\downarrow} +  n_{l+1,\uparrow}n_{l+1,\downarrow}) ]
\end{eqnarray}
and
\begin{equation}
{\cal H}_1 = -t(1-\delta)\sum_{\sigma}\sum_{l\in \rm o}
( c^\dagger_{l+1,\sigma}c_{l,\sigma} +  c^\dagger_{l,\sigma}c_{l+1,\sigma} ),
\end{equation}
where $\sum_{l \in \rm e(o)}$ refers to the summation over even (odd) $l$.  Since $|1-\delta|<<1$, ${\cal H}_0$ is  the unperturbed Hamiltonian, and ${\cal H}_1$ is the perturbation term.  The current operator is also rewritten as follows:
\begin{equation}
J= J_{0} + J_{1},
\end{equation}
where
\begin{equation}
J_{0} =  i t(1+\delta)\sum_{\sigma}\sum_{l\in \rm e}
( c^\dagger_{l,\sigma}c_{l+1,\sigma} -  c^\dagger_{l+1,\sigma}c_{l,\sigma} )
\end{equation}
and
\begin{equation}
J_{1} =  i t(1-\delta)\sum_{\sigma}\sum_{l\in \rm o}
( c^\dagger_{l,\sigma}c_{l+1,\sigma} -  c^\dagger_{l+1,\sigma}c_{l,\sigma} ).
\end{equation}

\subsection{Ground state}

Let us first consider the 0th-order ground state of the system.  In the completely decoupled case ($\delta=1$), the system is equivalent to $N_d(\equiv N/2)$ isolated dimers, each of which is described by the 2-site Hubbard model.  Thus all the eigenstates are represented as the direct products of the eigenstates in the 2-site Hubbard model,  whose Hamiltonian, eigenstates, and eigenvalues are summarized in Appendix.

The 0th-order ground state $|\psi^0_{0}\rangle$ is the direct product given by
\begin{equation}
|\psi^0_{0}\rangle = |G\rangle_{0} \otimes |G\rangle_{1} \otimes \cdots \otimes |G\rangle_{N_d-1},
\end{equation}
where $|G\rangle_{n}$ denotes the ground state of the $n$-th dimer ($n=0,1,\cdots,N_d-1$).  Note on the properties of $|G\rangle_n$ in the $U/t\to\infty$ limit that, as shown in Appendix, $|G\rangle_n$ is equivalent to the singlet dimer state of two S=1/2 quantum spins for $U/t\to\infty$.  Thus $|\psi^0_0\rangle$ in this limit is the product of the singlet dimers.  In our case with finite $U$, $|\psi^0_0\rangle$ contains a small component of doubly occupied site.

The 0th-order ground state energy is given by
\begin{eqnarray}
E^0_0&=&\frac{N}{2} \epsilon^{-}[t(1+\delta)] \nonumber\\
&=& \frac{N}{2} \left[ U/2 - \sqrt{U^2/4 + 4t^2(1+\delta)^2 }\right],
\label{eq:0thgene}
\end{eqnarray}
where $\epsilon^{-}(t)$ is the ground state energy of the 2-site Hubbard model with a transfer integral $t$.  It should be noted that the energy~(\ref{eq:0thgene}) is exact to the first order in ${\cal H}_1$ because of
\begin{equation}
\langle \psi^0_0|{\cal H}_1|\psi^0_0\rangle=0.
\end{equation}

\subsection{Intradimer CT state}

One class of relevant photoexcited states in the decoupling limit consists of intradimer charge-transfer (CT) states, which are obtained by operating $J_0$ onto $|\psi^0_0\rangle$.  We here introduce the following basis set:
\begin{eqnarray}
||O^2\rangle\rangle_{n} &\equiv& |G\rangle_{0} \otimes  \cdots \otimes |G\rangle_{n-1}\otimes |O^2\rangle_{n} \nonumber \\
&\otimes&  |G\rangle_{n+1} \otimes  \cdots \otimes |G\rangle_{N_d-1},
\end{eqnarray}
where $|O^2\rangle_n$ is the intradimer CT state of dimer $n$ [see Eq.~(\ref{eq:opt_dimer})].  Then we obtain
\begin{equation}
J_{\rm 0}|\psi^0_0\rangle = 2it(1+\delta)\alpha 
\sum_{n=0}^{N_d-1} ||O^2\rangle\rangle_{n}.
\label{eq:intraCT}
\end{equation}
The right-hand side of Eq.~(\ref{eq:intraCT}) is a linear combination of intradimer CT excitations $||O^2\rangle\rangle_n$, and all the states are degenerate eigenstates of ${\cal H}_0$ with energy
\begin{equation}
E^0_{\rm intra}=\left( \frac{N}{2}-1\right) \epsilon^{-}[t(1+\delta)] + U.
\end{equation}
Then its excitation energy is given by
\begin{eqnarray}
\Delta^0_{\rm intra} &\equiv& E^0_{\rm intra} - E^0_0  \nonumber \\
&=& U/2 + \sqrt{U^2/4 + 4t^2(1+\delta)^2 }.
\label{eq:gap_intra}
\end{eqnarray}
We note that these formulae for $E^0_{\rm intra}$ and $\Delta^0_{\rm intra}$ are also exact to the 1st order in ${\cal H}_1$ because of
\begin{equation}
\langle\langle O^2||_n{\cal H}_1||O^2\rangle\rangle_m =0 
\label{eq:1stintraCT}
\end{equation}
for any $n$ and $m$.  Equation~(\ref{eq:1stintraCT}) implies that the local CT excitation $|O^2\rangle_n$ transfers to the nearest-neighbor dimers $(n\pm 1)$ by at least 2-nd order processes of ${\cal H}_1$.

\subsection{Interdimer CT excitations}

\subsubsection{zeroth order in ${\cal H}_1$}

The other class of important photoexcited states consists of interdimer CT excitations.  Before discussing the details of these states, we introduce another useful basis.  Let us consider a state where the $n_1$-th dimer is in $|X^l_{\sigma}\rangle$ and the $n_2$-th ($n_1<n_2$) dimer is in  $|Y^{4-l}_{\bar{\sigma}}\rangle$.
Both $|X^l_{\sigma}\rangle$ and $|Y^{4-l}_{\bar{\sigma}}\rangle$ are eigenstates of the 2-site Hubbard model and the superscripts, $(l,4-l)=(1,3)$ or $(3,1)$, denote the number of electrons in each dimer.  The spin variable $\sigma$ takes $\uparrow$ or $\downarrow$, and $\bar{\sigma}$ denotes the opposite direction of $\sigma$.  We assume that the other dimers are in the ground state $|G\rangle$.  Such a state is represented by
\begin{eqnarray}
|X^l_{\sigma} Y^{4-l}_{\bar{\sigma}} \rangle_{n_1,n_2} &\equiv&
|G\rangle_{0} \otimes  \cdots   \otimes |G\rangle_{n_1-1} \otimes |X^l_{\sigma}\rangle_{n_1} \nonumber \\
&\otimes& |G\rangle_{n_1+1} \otimes  \cdots \otimes |G\rangle_{n_2-1} \nonumber \\
&\otimes& |Y^{4-l}_{\bar{\sigma}}\rangle_{n_2} \otimes |G\rangle_{n_2+1} \otimes  \cdots  \nonumber \\
&\otimes& |G\rangle_{N_d-1}.
\end{eqnarray}

The interdimer CT excitations are generated by operating $J_1$ onto $|\psi^0_0\rangle$.
Several terms thus created are classified into the lowest-energy terms and higher-energy terms as follows:
\begin{eqnarray}
J_{\rm 1}|\psi^0_0\rangle &=& it(1-\delta)\frac{(\alpha+\beta)^2}{4}\sum_{n=0}^{N_d-1}
(|O^3_{\uparrow}  E^1_{\downarrow}\rangle_{n,n+1} \nonumber \\
&+&|O^3_{\downarrow}  E^1_{\uparrow}\rangle_{n,n+1}  - |E^1_{\uparrow}  O^3_{\downarrow}\rangle_{n,n+1} \nonumber \\
&-&|E^1_{\downarrow} O^3_{\uparrow}\rangle_{n,n+1} ) \nonumber \\ 
&+& {\rm higher \ energy \ terms}.
\label{eq:interCT0}
\end{eqnarray}
All the states in the first part of Eq.~(\ref{eq:interCT0}) are degenerate eigenstates of ${\cal H}_0$ with energy
\begin{equation}
E^0_{\rm inter}=\left( \frac{N}{2}-2\right) \epsilon^{-}[t(1+\delta)] + U -2t(1+\delta).
\label{eq:ene_inter0}
\end{equation}
Its optical gap is given by
\begin{eqnarray}
\Delta^0_{\rm inter} &\equiv& E^0_{\rm inter} - E^0_0 \nonumber \\
&=& -2t(1+\delta) + 2\sqrt{U^2/4 + 4t^2(1+\delta)^2 }.
\label{eq:gap_inter0}
\end{eqnarray}

\begin{figure}[hbt]
\begin{center}
\includegraphics[width=7.0cm,clip]{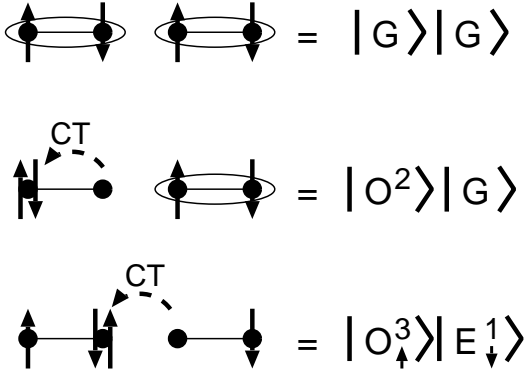}
\end{center}
\caption{Schematic picture of the ground state, the intradimer and the interdimer CT states. The ellipses show the singlet dimer states, and the up (down) arrows are electrons with up (down) spins.}
\label{fig:opt}
\end{figure}

It is obvious that $\Delta^0_{\rm intra}>\Delta^0_{\rm inter}$ for $U>0$.  Therefore the interdimer CT state is the lowest optical excitation.  This conclusion can be easily understood in a large $U$ system, which is appropriate to our study.  For $U/t>>1$, the excitation energies~(\ref{eq:gap_intra}) and (\ref{eq:gap_inter0}) are evaluated as
\begin{eqnarray}
\Delta^0_{\rm intra}&\sim&  U + 4 (1+\delta)^2 t^2/U, \label{eq:gap_intra0LU} \\
\Delta^0_{\rm inter}&\sim&  U -2t(1+\delta) + 8(1+\delta)^2t^2/U. \label{eq:gap_inter0LU}
\end{eqnarray}
The Coulomb term $U$ appears because of the creation of double occupation by the charge transfer (see Fig.~\ref{fig:opt}).
The last terms of Eqs.~(\ref{eq:gap_intra0LU}) and (\ref{eq:gap_inter0LU}) originate from the destruction of the singlet dimer(s).  The most important is the second term of Eq.~(\ref{eq:gap_inter0LU}): this is caused by the kinetic energy gain of the electron in  $|E^1_{\sigma}\rangle$ and that of the doubly occupied site in $|O^3_{\bar{\sigma}}\rangle$.  That is, in an interdimer CT state $|E^1_{\sigma}O^3_{\bar{\sigma}}\rangle$, the kinetic effect caused by moving the electron and the doubly occupied site reduces the excitation energy.

\begin{figure}[hbt]
\begin{center}
\includegraphics[width=7.0cm,clip]{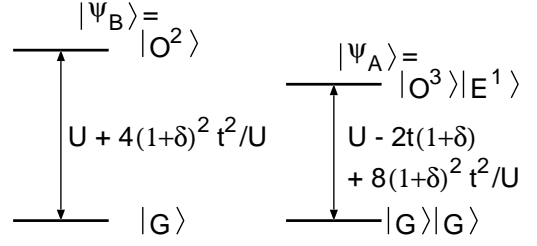}
\end{center}
\caption{Intradimer and interdimer CT states, and their excitation energies for $U/t>>1$. The spin indices in $|O^3_{\sigma}\rangle$ and in $|E^1_{\bar{\sigma}}\rangle$ are omitted.}
\label{fig:opt_0th}
\end{figure}

Now, it becomes clear which CT state corresponds to $|\psi_{\rm A(B)}\rangle$.  Because the energy of $|\psi_{\rm A}\rangle$ is lower than that of $|\psi_{\rm B}\rangle$, $|\psi_{\rm A}\rangle$ is the interdimer CT state and the $|\psi_{\rm B}\rangle$ is the intradimer CT state (see Fig.~\ref{fig:opt_0th}).  This conclusion is quite natural because the spectral intensity of $|\psi_{\rm B}\rangle$ becomes dominant while that of $|\psi_{\rm A}\rangle$ decreases as $\delta$ approaches 1.  In addition, our interpretation of $|\psi_{\rm B}\rangle$ is consistent with that in the preceding studies.~\cite{campbell,gallagher,lyo,gebhard1,gebhard2}

\subsubsection{First order in ${\cal H}_1$}

Here we discuss the 1st-order correction to the optical gap due to the interdimer CT state.  We have found that the 0th-order interdimer CT state~(\ref{eq:interCT0}) consists of a number of interdimer CT excitations $|O^3_{\sigma}E^1_{\bar{\sigma}}\rangle_{n,n+1}$ and $|E^1_{\sigma}O^3_{\bar{\sigma}}\rangle_{n,n+1}$ ($0\le n\le N_d$).
  All these CT states have degenerate energy~(\ref{eq:ene_inter0}).  In addition, states that contain $|O^3_{\sigma}\rangle$ and $|E^1_{\bar{\sigma}}\rangle$ in separate dimers $n_1,n_2$ ($|n_2-n_1|>1$) also have the same energy~(\ref{eq:ene_inter0}).
Thus we need to perform the 1st-order degenerate perturbation theory to obtain the energy gap to the 1st order in ${\cal H}_1$.  For this purpose we calculate the matrix elements of ${\cal H}_1$ for the degenerate states $|X^l_{\sigma}Y^{4-l}_{\bar{\sigma}}\rangle_{n_1,n_2}$, with $(X^l,Y^{4-l})=(O^3,E^1)$ or $(E^1,O^3)$, to obtain
\begin{eqnarray}
{\cal H}_1 |X^l_{\sigma} Y^{4-l}_{\bar{\sigma}}\rangle_{n_1,n_2} 
&=& \frac{t(1-\delta)}{4}(\alpha+\beta)^2 (
 |X^l_{\sigma} Y^{4-l}_{\bar{\sigma}}\rangle_{n_1,n_2+1} \nonumber \\
&+&  |X^l_{\sigma} Y^{4-l}_{\bar{\sigma}}\rangle_{n_1,n_2-1}
 +  |X^l_{\sigma} Y^{4-l}_{\bar{\sigma}}\rangle_{n_1+1,n_2} \nonumber \\
&+&  |X^l_{\sigma} Y^{4-l}_{\bar{\sigma}}\rangle_{n_1-1,n_2} 
) \nonumber \\
&&\quad {\rm for} \quad n_2-n_1>1,
\label{eq:inter_h1_1}
\end{eqnarray}
and
\begin{eqnarray}
{\cal H}_1 |X^l_{\sigma} Y^{4-l}_{\bar{\sigma}}\rangle_{n_1,n_2} &=& \frac{t(1-\delta)}{4}(\alpha+\beta)^2 (
 |X^l_{\sigma} Y^{4-l}_{\bar{\sigma}}\rangle_{n_1,n_2+1} \nonumber \\
&+&  |X^l_{\sigma} Y^{4-l}_{\bar{\sigma}}\rangle_{n_1-1,n_2} ) \nonumber \\
&&\quad {\rm for} \quad n_2-n_1=1.
\label{eq:inter_h1_2}
\end{eqnarray}
Here, other states with higher (or lower) energies are omitted because they are not needed at present.

\begin{figure}[hbt]
\begin{center}
\includegraphics[width=7.0cm,clip]{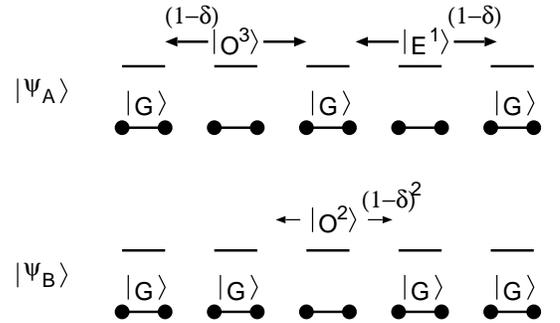}
\end{center}
\caption{Dynamics of excited dimers in the optically excited states.  Effective transfer integrals for $|O^3\rangle$ and $|E^1\rangle$, and that for $|O^2\rangle$ are very different as indicated.}
\label{fig:opt1}
\end{figure}

The relations~(\ref{eq:inter_h1_1}) and (\ref{eq:inter_h1_2}) show that the local excitations $|O^3_{\sigma}\rangle$ and $|E^1_{\bar{\sigma}}\rangle$ move in the system by the 1st-order process of ${\cal H}_1$, as illustrated in Fig.~\ref{fig:opt1}.  This is quite different from the intradimer CT state, where the local CT excitation $|O^2\rangle$ transfers by at least 2nd-order processes of ${\cal H}_1$.

The 1st-order Hamiltonian ${\cal H}_1$ defined by (\ref{eq:inter_h1_1}) and (\ref{eq:inter_h1_2}) can be diagonalized analytically in the subspace spanned by $|X^l_{\sigma} Y^{4-l}_{\bar{\sigma}}\rangle_{n_1,n_2}$.  We here introduce a new basis set:
\begin{eqnarray}
|n_1,n_2\rangle &\equiv& \frac{1}{2}
\left[ |O^3_{\uparrow} E^1_{\downarrow}\rangle_{n_1,n_2} + 
|O^3_{\downarrow} E^1_{\uparrow}\rangle_{n_1,n_2} \right. \nonumber \\
 && \left. - |E^1_{\uparrow} O^3_{\downarrow}\rangle_{n_1,n_2}
- |E^1_{\downarrow} O^3_{\uparrow}\rangle_{n_1,n_2} \right].
\end{eqnarray}
By using this basis set, $J_{\rm 1}|\psi^0_0\rangle$ is written as 
\begin{eqnarray}
J_{\rm 1}|\psi^0_0\rangle &=& it(1-\delta)\frac{(\alpha+\beta)^2}{2}\sum_{n=0}^{N_d-1}|n,n+1\rangle,
\label{eq:interCT0_basis}
\end{eqnarray}
where the higher-energy terms are omitted.
Then we obtain
\begin{eqnarray}
{\cal H}_1|n_1,n_2\rangle &=& \frac{t(1-\delta)}{4}(\alpha+\beta)^2
( |n_1+1,n_2\rangle + |n_1-1,n_2\rangle \nonumber \\
&&+ |n_1,n_2+1\rangle + |n_1,n_2-1\rangle) \nonumber \\
&&  {\rm for \ n_2-n_1\ge 2,}
\end{eqnarray}
and 
\begin{eqnarray}
{\cal H}_1|n_1,n_2\rangle &=& \frac{t(1-\delta)}{4}(\alpha+\beta)^2
(  |n_1-1,n_2\rangle + |n_1,n_2+1\rangle ) \nonumber \\ 
&& {\rm for \ n_2-n_1 = 1}.
\end{eqnarray}
It is noted that the following boundary condition should be used:
\begin{eqnarray}
|n_1,n_2=N_d\rangle  &=& |0,n_1\rangle \nonumber \\
|n_1=-1,n_2\rangle &=& |n_2,N_d-1\rangle.
\end{eqnarray}
Then this problem is reduced to the spinless free-fermion system with two particles under 
the antiperiodic boundary condition. The eigenstates are written as
\begin{eqnarray}
|\psi(k_1,k_2)\rangle &=& 1/N_d \sum_{0\le n_1 < n_2 \le N_d-1} |n_1,n_2\rangle \nonumber \\
&\times& [e^{i(k_1n_1+k_2n_2) }- e^{i(k_2n_1+k_1n_2)} ],
\label{eq:psi_A1}
\end{eqnarray}
where
\begin{equation}
k_{1(2)}=\frac{2m_{1(2)}+1}{N_d}\pi \quad (m_{1(2)}=0,1,\cdots,N_d-1),
\end{equation}
with $k_1\ne k_2$. Their energies are given by
\begin{equation}
E^1(k_1,k_2)=\frac{t(1-\delta)}{2}(\alpha+\beta)^2 [ \cos(k_1)+\cos(k_2)].
\end{equation}

Since $|\psi^0_{\rm A}\rangle$ is the lowest-energy state of Eq.~(\ref{eq:psi_A1}) 
with the total wave number being zero, this state is given by
\begin{equation}
|\psi^0_{\rm A}\rangle = |\psi(k,-k)\rangle \quad {\rm with } \quad  k=\frac{N-2}{N}\pi.
\end{equation}
For $N=8$, the wave numbers are $k_1=3/4\pi, k_2=-3/4\pi$, and then we have
\begin{equation}
E^1_{\rm inter}(k_1,k_2)=-\frac{t(1-\delta)}{\sqrt{2}}(\alpha+\beta)^2.
\end{equation}
Thus the optical gap to the 1st order is
\begin{equation}
\Delta^1_{\rm inter} \equiv E^0_{\rm inter}+E^1_{\rm inter} - E^0_0.
\end{equation}

\begin{figure}[hbt]
\begin{center}
\includegraphics[width=7.0cm,clip]{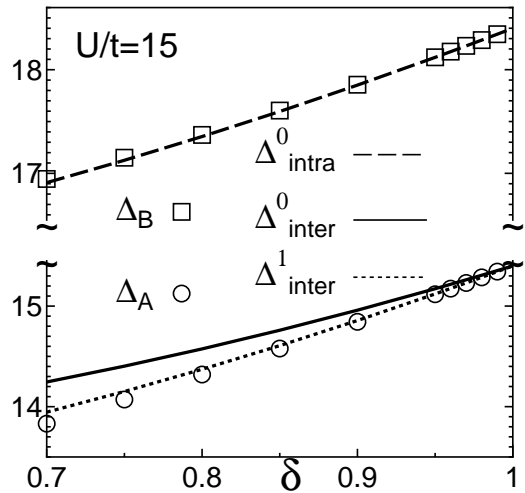}
\end{center}
\caption{Optical excitation energies.  The symbols show the exact-diagonalization results, and the lines the perturbative results.}
\label{fig:optgap}
\end{figure}

In Fig.~\ref{fig:optgap}, the perturbative results for the optical excitation energies $\Delta^0_{\rm intra}$,  $\Delta^0_{\rm inter}$, and $\Delta^1_{\rm intra}$ are compared with the exact results obtained by the numerical diagonalization.
The numerically calculated, optical excitation energies of $|\psi_{\rm A(B)}\rangle$ are denoted by $\Delta_{\rm A(B)}$.
  The analytical results for $\Delta^0_{\rm intra}$ show good agreement with the numerical results for $\Delta_{\rm B}$.  Hence we confirm that the state $|\psi_{\rm B}\rangle$ is the intradimer CT state.
The results for $\Delta^0_{\rm inter}$ agree with the numerical results for $\Delta_{\rm A}$ around $\delta\sim 1$.  As $\delta$ decreases, the difference between them becomes relatively large compared to that between $\Delta^0_{\rm intra}$ and $\Delta_{\rm B}$ because of the finite 1st-order correction due to the dynamics of $|O^3_{\sigma}\rangle$ and $|E^1_{\bar{\sigma}}\rangle$.
The agreement between $\Delta^1_{\rm inter}$ and the diagonalization results supports our argument on the 1st-order perturbation.

\subsubsection{Origin of the ``inverse'' spin-Peierls instability}

\begin{figure}[hbt]
\begin{center}
\includegraphics[width=7.0cm,clip]{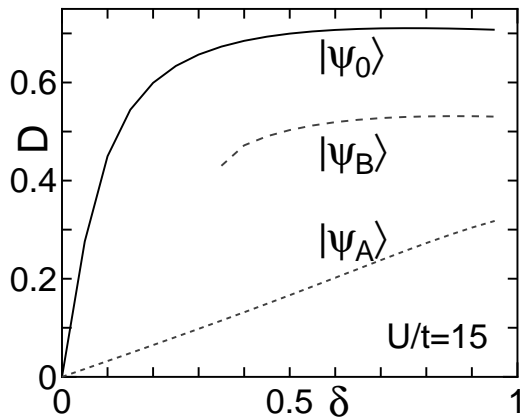}
\end{center}
\caption{Spin dimerization in the three states.}
\label{fig:D}
\end{figure}

Up to now, we have clarified how the CT occurs in the relevant photoexcited states in the dimerized phase.  Here, we focus on the relation between the CT type and the (in)stability against the lattice dimerization.  The dimerization in the spin-Peierls system is caused by the energy gain due to the spin singlet states formed on the dimers.  Hence whether the lattice dimerization is stable or not in the CT states depends on how the spin singlet states are destroyed by the CT excitations.  To examine this, we calculated a quantity that measures the strength of the spin dimerization:
\begin{equation}
 D\equiv | \langle \vec{S}_0\cdot\vec{S}_1 \rangle - \langle \vec{S}_1\cdot\vec{S}_2 \rangle|.
\end{equation}
The obtained results are shown in Fig.~\ref{fig:D}.  The ground state has large spin dimerization even when $\delta$ is not so large: at $\delta=0.2$, $D$ is about 80$\%$ of its maximum value.
By contrast, that of $|\psi_{\rm A}\rangle$ is strongly reduced, which indicates that a large part of the spin dimers is destroyed by the interdimer CT excitation.  In $|\psi_{\rm B}\rangle$, $D$ has a larger value than that of $|\psi_{\rm A}\rangle$ although it is of course smaller than that of $|G\rangle$.  This result is consistent with the fact that the strength of $|\psi_{\rm A}\rangle$ for destabilizing the dimer phase is much larger than that of $|\psi_{\rm B}\rangle$.

\section{midgap states in $\sigma_1(\omega)$}

\subsection{Numerical results}

Here, we discuss optical responses of the system after the photoexcitation.  Our interest in this aspect has been stimulated by the experiment by Okamoto {\it et al.}, who have reported the appearance of a midgap state in the reflectivity spectrum of K-TCNQ immediately after the photoexcitation.~\cite{KTCNQ2}  Because the midgap state is observed after the photoexcitation causing the inverse spin-Peierls transition, the realized state is assumed to be $|\psi_{\rm A}\rangle$.  In addition, we note that it is unnecessary to take account of the lattice relaxation because the midgap state already appears before the lattice starts to move.~\cite{KTCNQ2}  Then the key quantity is the optical conductivity in the state $|\psi_{\rm A}\rangle$ given by
\begin{equation}
\sigma_1(\omega)\equiv D_1\delta(\omega)  +\sigma_1^{\rm reg}(\omega)  +\sigma_1^{\rm reg}{'}(\omega),
\label{eq:sigma1}
\end{equation}
with
\begin{eqnarray}
&&\sigma_1^{\rm reg}(\omega) \equiv    \nonumber \\
 &-&\frac{1}{N\omega}{\rm Im}\left[\langle\psi_{\rm A}|J\frac{1}{\omega+i\epsilon+E_{\rm A} -{\cal H}}J|\psi_{\rm A}\rangle \right], \ \ \ \ \ \ 
\end{eqnarray}
and
\begin{eqnarray}
&&\sigma_1^{\rm reg}{'}(\omega)  \equiv \ \ \ \ \ \ \ \ \ \nonumber \\
& &\frac{1}{N\omega}{\rm Im}\left[\langle\psi_{\rm A}|J\frac{1}{\omega+i\epsilon-E_{\rm A} +{\cal H}}J|\psi_{\rm A}\rangle\right].\ \ \ \ \ 
\end{eqnarray}
Here, $D_1$ is the Drude weight of $|\psi_{\rm A}\rangle$ defined by
\begin{equation}
D_1 = - \frac{\pi}{N} \langle\psi_{\rm A}|K|\psi_{\rm A}\rangle - \frac{2\pi}{N}\sum_{n\ne \rm A}\frac{|\langle \psi_{ n}|J|\psi_{\rm A}\rangle|^2}{E_{n}-E_{\rm A}},
\end{equation}
where $E_{\rm A}$ is the energy of $|\psi_{\rm A}\rangle$.~\cite{maeshima1,maeshima2}

\begin{figure}[hbt]
\begin{center}
\includegraphics[width=7.0cm,clip]{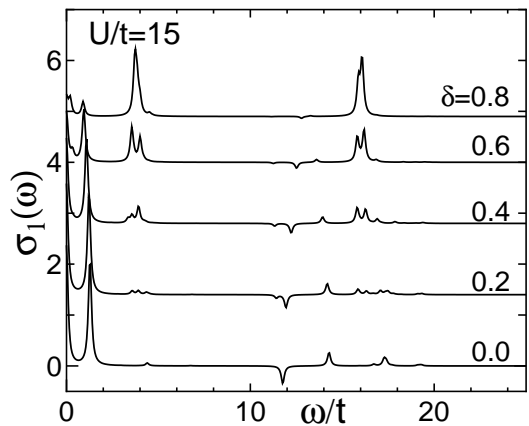}
\end{center}
\caption{Optical conductivity spectra in the photoexcited state $|\psi_{\rm A}\rangle$ for $U/t=15$.}
\label{fig:sw1.1}
\end{figure}

Figure~\ref{fig:sw1.1} shows numerical diagonalization results of $\sigma_1(\omega)$ for $U/t=15$ and $N=8$.  At $\delta=0$, there appear the large Drude peak and another pronounced low-energy peak.  The presence of the Drude peak suggests the metallic property caused by the photodoped carriers, the holon and the doublon.~\cite{maeshima1,maeshima2}   The latter low-energy peak corresponds to the excitation from $|\psi_{\rm A}\rangle$ to the nearly degenerate excited state (named $|\psi_{\rm A'}\rangle$),~\cite{maeshima1,maeshima2} which is an important feature of 1D uniform Mott insulators.~\cite{mizuno}
As $\delta$ increases, these peaks lose their intensity and several new peaks appear.  The high-energy peak around $\omega\sim U$ corresponds to the intradimer CT excitation from $|G\rangle$ to $|O^2\rangle$, which is also observed in $\sigma(\omega)$.  Since $|\psi_{\rm A}\rangle$ still includes a number of dimers in the state $|G\rangle$ except for the charge-transfered dimers $|E^1_{\sigma} O^3_{\bar{\sigma}}\rangle$, this peak appears.  More interesting are midgap peaks around $\omega\sim 4t$.  The midgap peaks develop with increasing $\delta$, indicating that they are closely related to the dimerization.  
At first glance, the midgap state seems to belong to a different class from the nearly degenerate excited state $|\psi_{\rm A'}\rangle$.  The midgap state appears only for finite $\delta$ while  $|\psi_{\rm A'}\rangle$ exists even for $\delta=0$.  In addition, the energy of the midgap state ($\omega/t \sim 4$) is far from that of $|\psi_{\rm A'}\rangle$ ($\omega/t \sim 1$).  Since $|\psi_{\rm A'}\rangle$ is close to $|\psi_{\rm A}\rangle$ even for finite $\delta$, $|\psi_{\rm A'}\rangle$ would be a linear combination of $|E^1_{\sigma}O^3_{\bar{\sigma}}\rangle$, which are also the constituents of $|\psi_{\rm A}\rangle$.  By contrast, the midgap state consists of other types of excitations, as discussed in the next subsection.

\subsection{Zeroth order in ${\cal H}_1$}

As a starting point to discuss the origin of the midgap states, let us present again the analytical form of $|\psi_{\rm A}\rangle$: the perturbative result of the wave function of $|\psi_{\rm A}\rangle$ is given by Eq.~(\ref{eq:psi_A1}) with $k_1=-k_2=k$,
\begin{eqnarray}
|\psi^0_{\rm A}\rangle&=&
1/N_d\sum_{n_1 < n_2} [e^{ik(n_1-n_2) }- e^{-ik(n_1-n_2)} ] \nonumber \\
&\times&\frac{1}{2}
   (|O^3_{\uparrow} E^1_{\downarrow}\rangle_{n_1,n_2} + |O^3_{\downarrow} E^1_{\uparrow}\rangle_{n_1,n_2} \nonumber \\
&-&  |E^1_{\uparrow} O^3_{\downarrow}\rangle_{n_1,n_2} - |E^1_{\downarrow} O^3_{\uparrow}\rangle_{n_1,n_2}),
\end{eqnarray}
with $k=\pi(N-2)/N$.
Since the spectral intensity of the midgap states increases with $\delta$, candidates for the midgap states would be the intradimer CT excitations obtained by operating $J_0$ onto $|\psi^0_{\rm A}\rangle$ as follows:
\begin{eqnarray}
J_0|\psi^0_{\rm A}\rangle&=&
\frac{it(1+\delta)}{N_d} \sum_{n_1 < n_2} [e^{ik(n_1-n_2) }- e^{-ik(n_1-n_2)} ] \nonumber \\
&\times&\frac{1}{2}
   (|E^3_{\uparrow} E^1_{\downarrow}\rangle_{n_1,n_2} + |E^3_{\downarrow} E^1_{\uparrow}\rangle_{n_1,n_2} \nonumber \\
&+&  |O^3_{\uparrow} O^1_{\downarrow}\rangle_{n_1,n_2} + |O^3_{\downarrow} O^1_{\uparrow}\rangle_{n_1,n_2} \nonumber \\
&-&  |E^1_{\uparrow} E^3_{\downarrow}\rangle_{n_1,n_2} - |E^1_{\downarrow} E^3_{\uparrow}\rangle_{n_1,n_2} \nonumber \\
&-&|O^1_{\uparrow} O^3_{\downarrow}\rangle_{n_1,n_2} - |O^1_{\downarrow} O^3_{\uparrow}\rangle_{n_1,n_2}),
\label{eq:opt_mid}
\end{eqnarray}
where other states with higher energy are omitted.  All the states in the above equation, $|E^l_{\sigma} E^{(4-l)}_{\bar{\sigma}}\rangle$ and  $|O^l_{\sigma} O^{(4-l)}_{\bar{\sigma}}\rangle$, are degenerate eigenstates of ${\cal H}_0$ with energy,
\begin{eqnarray}
E^0_{\rm mid1}
&=& \frac{N}{2} \left[ U/2 - \sqrt{U^2/4 + 4t^2(1+\delta)^2 }\right] \nonumber \\
&&+ 2\sqrt{U^2/4 + 4t^2(1+\delta)^2 }.
\end{eqnarray}

\begin{figure}[hbt]
\begin{center}
\includegraphics[width=7.4cm,clip]{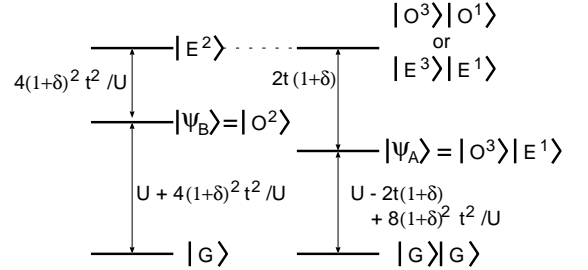}
\end{center}
\caption{Midgap and other relevant states. The energy differences are also shown for $U/t>>1$.}
\label{fig:opt3}
\end{figure}

In addition, we find that another class of states contribute to the midgap states.  Operating $J_1$ onto $|\psi^0_{\rm A}\rangle$, we obtain
\begin{equation}
J_1|\psi^0_{\rm A}\rangle
= - \frac{2t(1-\delta)}{N_d}(\alpha^2-\beta^2) \sin(k) \sum_n ||E^2\rangle\rangle_n,
\end{equation}
where the new state $||E^2\rangle\rangle_n$ is defined by
\begin{eqnarray}
||E^2\rangle\rangle_{n} &\equiv& |G\rangle_{0} \otimes  \cdots \otimes |G\rangle_{n-1}\otimes |E^2\rangle_{n} \nonumber \\
&\otimes&  |G\rangle_{n+1} \otimes  \cdots \otimes |G\rangle_{N_d-1}.
\end{eqnarray}
$||E^2\rangle\rangle_n$ is also an degenerate eigenstate of ${\cal H}_0$ with energy,
\begin{eqnarray}
E^0_{\rm mid2} 
&=& \frac{N}{2} \left[ U/2 - \sqrt{U^2/4 + 4t^2(1+\delta)^2 }\right] \nonumber \\
&&+2\sqrt{U^2/4 + 4t^2(1+\delta)^2 } \\
&=&E^0_{\rm mid1}.
\end{eqnarray}
Figure~\ref{fig:opt3} displays the relevant states and their relative energies given by 
\begin{eqnarray}
 E^0_{\rm mid1} - E^0_{\rm inter} &=& 2t(1+\delta),\label{eq:midgap_gap} \\
 E^0_{\rm mid2} - E^0_{\rm intra} &=& \epsilon^+[t(1+\delta)]-U\nonumber \\ 
                                  &\sim& 4(1+\delta)^2t^2/U \nonumber \\
 && {\rm for }\quad U/t>>1.
\end{eqnarray}
Here it should be noted that the energy difference~(\ref{eq:midgap_gap}) between $|\psi^0_{\rm A}\rangle$ and the excited state $J_0|\psi^0_{\rm A}\rangle$ is quite close to the gap ($\sim 4t$) observed in Fig.~\ref{fig:sw1.1}.  Thus these excited states are considered to be the midgap states.

\subsection{First order in ${\cal H}_1$}

\begin{figure}[hbt]
\begin{center}
\includegraphics[width=7.0cm,clip]{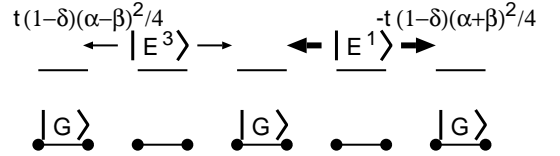}
\end{center}
\caption{Dynamics of excited dimers in the midgap state.  It contains a light ``particle'' $|E^1\rangle$ and a heavy ``particle'' $|E^3\rangle$.}
\label{fig:midgap-opt2}
\end{figure}

Without perturbation, the midgap states are degenerate.
However, the numerical results in Fig.~\ref{fig:sw1.1} show that these states are split for finite $(1-\delta)$ by the perturbation.
Then we discuss the effect of the interdimer interaction ${\cal H}_1$ on the midgap states to the 1st order.
The following degenerate states should be taken into account:
$|E^3_{\sigma} E^1_{\bar{\sigma}}\rangle_{n_1,n_2}$,
$|O^3_{\sigma} O^1_{\bar{\sigma}}\rangle_{n_1,n_2}$, and $||E^2\rangle\rangle_n$.
For simplicity,  we introduce a basis set given by
\begin{eqnarray}
|\overline{n_1,n_2}\rangle &\equiv& \frac{1}{2}
(|E^3_{\uparrow}E^1_{\downarrow}\rangle_{n_1,n_2} + |E^3_{\downarrow}E^1_{\uparrow}\rangle_{n_1,n_2} \nonumber \\
&-& |O^1_{\uparrow}O^3_{\downarrow}\rangle_{n_1,n_2} - |O^1_{\downarrow}O^3_{\uparrow}\rangle_{n_1,n_2} )
\quad {\rm for}\quad n_1<n_2\nonumber\\
\nonumber\\
&\equiv& \frac{1}{2}
(|E^1_{\uparrow}E^3_{\downarrow}\rangle_{n_2,n_1} + |E^1_{\downarrow}E^3_{\uparrow}\rangle_{n_2,n_1} \nonumber \\
&-& |O^3_{\uparrow}O^1_{\downarrow}\rangle_{n_2,n_1} - |O^3_{\downarrow}O^1_{\uparrow}\rangle_{n_2,n_1} )
\quad {\rm for} \quad n_2<n_1. \nonumber \\
\end{eqnarray}
By using this, we represent the relation~(\ref{eq:opt_mid}) in a simpler form:
\begin{eqnarray}
J_0|\psi^0_{\rm A}\rangle&=&
\frac{it(1+\delta)}{N_d} \sum_{n_1 < n_2} [e^{ik(n_1-n_2) }- e^{-ik(n_1-n_2)} ] \nonumber \\
&\times&(|\overline{n_1,n_2}\rangle - |\overline{n_2,n_1}\rangle).
\end{eqnarray}
The matrix elements of ${\cal H}_1$ are calculated as follows:
\begin{eqnarray}
{\cal H}_1|\overline{n_1,n_2}\rangle 
&=& \frac{t(1-\delta)}{4}(\alpha+\beta)^2 (|\overline{n_1,n_2+1}\rangle + |\overline{n_1,n_2-1}\rangle) \nonumber \\
&-& \frac{t(1-\delta)}{4}(\alpha-\beta)^2 (|\overline{n_1+1,n_2}\rangle + |\overline{n_1-1,n_2}\rangle) \nonumber \\
\label{eq:midgap_11}
\end{eqnarray}
for $n_2-n_1 >1$,
\begin{eqnarray}
{\cal H}_1|\overline{n_1,n_2}\rangle 
&=& \frac{t(1-\delta)}{4}(\alpha+\beta)^2 |\overline{n_1,n_2+1}\rangle \nonumber \\ 
&-& \frac{t(1-\delta)}{4}(\alpha-\beta)^2 |\overline{n_1-1,n_2}\rangle) \nonumber \\ 
&-& \frac{t(1-\delta)}{2}(\alpha+\beta)^2 ||E^2\rangle\rangle_{n_1} \nonumber \\ 
&+& \frac{t(1-\delta)}{2}(\alpha-\beta)^2 ||E^2\rangle\rangle_{n_2} 
\label{eq:midgap_12}
\end{eqnarray}
for $n_2 = n_1 + 1$,
\begin{eqnarray}
{\cal H}_1|n_1,n_2\rangle 
&=& \frac{t(1-\delta)}{4}(\alpha+\beta)^2 |\overline{n_1,n_2-1}\rangle  \nonumber \\ 
&-& \frac{t(1-\delta)}{4}(\alpha-\beta)^2 |\overline{n_1+1,n_2}\rangle) \nonumber \\ 
&+& \frac{t(1-\delta)}{2}(\alpha+\beta)^2 ||E^2\rangle\rangle_{n_1} \nonumber \\ 
&-& \frac{t(1-\delta)}{2}(\alpha-\beta)^2 ||E^2\rangle\rangle_{n_2}
\label{eq:midgap_13}
\end{eqnarray}
for $n_2 = n_1 - 1$, and
\begin{eqnarray}
{\cal H}_1||E^2\rangle\rangle_{n}
&=& -\frac{t(1-\delta)}{2}(\alpha+\beta)^2 (|\overline{n,n+1}\rangle - |\overline{n,n-1}\rangle) \nonumber \\
&-& \frac{t(1-\delta)}{4}(\alpha-\beta)^2 (|\overline{n+1,n}\rangle - |\overline{n-1,n}\rangle). \nonumber \\
\label{eq:midgap_14}
\end{eqnarray}
Equations~(\ref{eq:midgap_11}),~(\ref{eq:midgap_12}),~(\ref{eq:midgap_13}), and (\ref{eq:midgap_14}) show several characteristics of the local excitations included in the midgap states.
For example, 
the local states $|E^1_{\sigma}\rangle$ and $|O^3_{\tau}\rangle$ move in the system with the larger transfer integral $-t(1-\delta)(\alpha+\beta)^2/4$  while the states $|O^1_{\sigma}\rangle$ and $|E^3_{\tau}\rangle$ move with the smaller transfer integral $t(1-\delta)(\alpha-\beta)^2/4$ (see Fig.~\ref{fig:midgap-opt2}).
In particular, if the terms containing $||E^2\rangle\rangle$ are omitted, the system becomes equivalent to the spinless free-fermion system with a light ``particle'' ($|E^1_{\sigma}\rangle$ or $|O^3_{\tau}\rangle$) and a heavy ``particle'' ($|O^1_{\sigma}\rangle$ or $|E^3_{\tau}\rangle$).
The terms containing $||E^2\rangle\rangle$ mean that a collision between the light ``particle'' and the heavy ``particle'' creates $||E^2\rangle\rangle$. 
It is also found that the state $||E^2\rangle\rangle$ splits into a pair of the light ``particle'' and the heavy ``particle''.
Here we note that the following boundary conditions should be used:
\begin{eqnarray}
|n_1,n_2=N_d\rangle  &=& -|n_1,0\rangle \nonumber \\
|n_1,n_2=-1\rangle &=& -|n_1,N_d-1\rangle \nonumber\\
|n_1=N_d,n_2\rangle  &=& -|0,n_2\rangle \nonumber \\
|n_1=-1,n_2\rangle &=& -|N_d-1,n_2\rangle \nonumber \\
||E^2\rangle\rangle_{N_d} &=& ||E^2\rangle\rangle_{0} \nonumber \\
||E^2\rangle\rangle_{-1} &=& ||E^2\rangle\rangle_{N_d-1}.
\end{eqnarray}

The effective Hamiltonian derived here is difficult to diagonalize analytically.
Hence we carry out the numerical diagonalization to obtain the $i$-th eigenstate in the form:
\begin{eqnarray}
 |\psi^i_{{\rm mid}}\rangle &=& \sum_{0\le n_1<n_2\le N_d-1} a^i(n_1,n_2)|\overline{n_1,n_2}\rangle \nonumber \\
&+& \sum_{0\le n_2<n_1\le N_d-1} \bar{a}^i(n_1,n_2)|\overline{n_1,n_2}\rangle \nonumber \\
&+& \sum_{0\le n\le N_d-1} b^i(n)||E^2\rangle\rangle_n.
\end{eqnarray}
To pick up optically allowed excited states, we calculate the spectral intensity to the 0th order in ($1-\delta$):
\begin{equation}
 I^0_i \equiv |\langle \psi^0_{\rm A}|J_0|\psi^i_{{\rm mid}} \rangle |^2/t^2,\label{eq:intensity}
\end{equation}
where
\begin{eqnarray}
\langle \psi^0_{\rm A}|J_0|\psi^i_{{\rm mid}}\rangle &=& \frac{2t(1+\delta)}{N_d}\nonumber\\
&\times& \{ \sum_{n_1<n_2} \sin [k(n_2-n_1)] a^i(n_1,n_2) \nonumber \\
&+& \sum_{n_2<n_1} \sin [k(n_2-n_1)] \bar{a}^i(n_1,n_2) \}.
\label{eq:intensity2}
\end{eqnarray}

\begin{figure}[hbt]
\begin{center}
\includegraphics[width=7.0cm,clip]{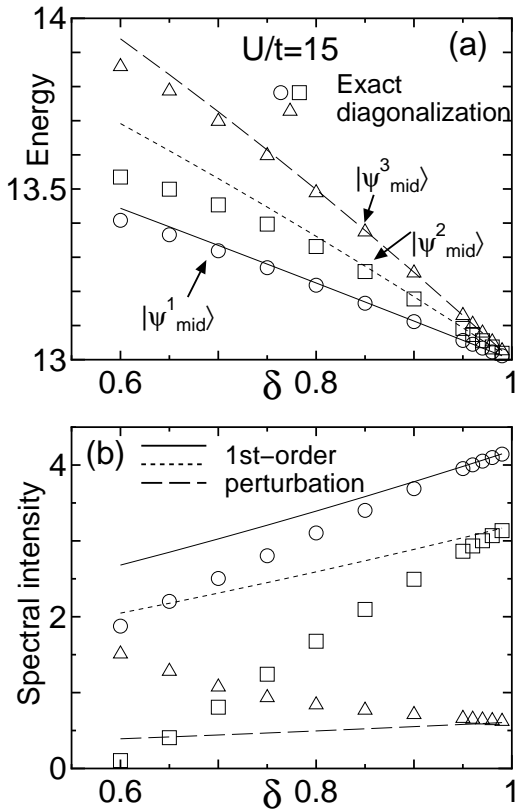}
\end{center}
\caption{Calculated energies and spectral intensities for the midgap states.  The symbols show the results of the exact diagonalization of the Hamiltonian~(\ref{eq:ham}) and the lines denote the results of the 1st-order perturbation theory.  The circles and the solid lines are for $|\psi^1_{\rm mid}\rangle$, the squares and the dotted lines for $|\psi^2_{\rm mid}\rangle$, and the triangles and the dashed lines for $|\psi^3_{\rm mid}\rangle$.}
\label{fig:midgap-E-diff2}
\end{figure}

In Fig.~\ref{fig:midgap-E-diff2}, the perturbative results for the energies and the spectral intensities are compared with the exact-diagonalization results for $N=8$.  The numerical results show that there are three photoexcited states, which are well reproduced by our perturbation theory.  We obtain two isolated states, $|\psi^1_{\rm mid}\rangle$ and $|\psi^3_{\rm mid}\rangle$, and several degenerate states that are collectively named $|\psi^2_{\rm mid}\rangle$.  The spectral intensity of  $|\psi^2_{\rm mid}\rangle$ is obtained by the summation of those of the degenerate states.
The higher-order effect would lift the degeneracy for $|\psi^2_{\rm mid}\rangle$. Some of thus split states must be optically active.  As for the spectral intensity, the perturbative results are good only in the vicinity of  $\delta=1$.  This is because the obtained wave function is correct only to the $0$-th order in $1-\delta$ and also because the interdimer current operator $J_1$ is omitted in Eq.~(\ref{eq:intensity}).

\section{summary and discussion}

\begin{figure}[hbt]
\begin{center}
\includegraphics[width=7.0cm,clip]{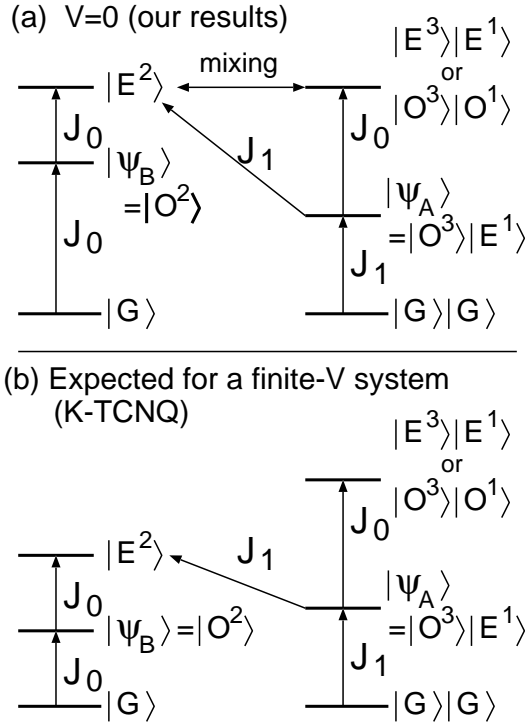}
\end{center}
\caption{Schematic picture of the relevant optical excitations for (a) $V=0$ and (b) finite $V$.}
\label{fig:opt_for_ktcnq}
\end{figure}

We have studied the properties of the photoexcited states in the 1D dimerized Hubbard model coupled with the lattice.
A notable point of our conclusion is that the lowest photoexcited state in this model is the interdimer CT state.
It is true that this point is inconsistent with the experimental result for K-TCNQ, where the lowest-energy peak in  $\sigma(\omega)$ of K-TCNQ has dominant spectral intensity.~\cite{yakushi,okamoto,yamaguchi}
This experimental fact implies that the lowest photoexcited state is an intradimer CT state, as the authors pointed out.~\cite{okamoto,yamaguchi}
This discrepancy would be due to the finite nearest-neighbor Coulomb interaction, often denoted by $V$, of K-TCNQ.  The intradimer CT state could lower its energy by the effective attractive interaction $V$ between the holon and the doublon.
This is because, in the intradimer CT state, the mean distance between them is smaller than that in the interdimer CT state.  The expected energy diagram for a finite-$V$ system is compared with our results for $V=0$ in Fig.~\ref{fig:opt_for_ktcnq}. 

However, in the experiments of the photoinduced inverse spin-Peierls transition,  the excitation energy so far corresponds to the energy of the interdimer CT state rather than that of the lowest-energy intradimer CT state.~\cite{KTCNQ1,KTCNQ2}  In this sense, the experimental result is consistent with our conclusion that the interdimer CT state destabilizes the dimerized phase, leading to the inverse spin-Peierls transition.  
Our assumption that the realized state after the photoexcitation is the interdimer CT state is justified in discussing the photoinduced dynamics observed so far.
If an experiment were performed in which the energy of the pulsed laser is lowered to coincide with the excitation energy of the lowest-energy intradimer CT state, the dimerization would be only weakly reduced and the following lattice dynamics would reflect the local character of the intradimer CT state, which is quite different from the character of the interdimer CT states.


The photoconductivity experiment of K-TCNQ also supports our conclusion on the photoexcited state.
The experimental results shown in Fig.2 (a) of Ref.~\onlinecite{KTCNQ2} reveal the conducting nature of the photoexcited state.
With the free-carrier-like charged dimer states $|E^1\rangle$ and $|O^3\rangle$, 
the interdimer CT state is the best candidate for the photoexcited state.
By contrast the intradimer CT state has an exciton-like bound state that does not have the conducting nature.  This situation would remain unchanged for finite $V$ systems.


Here we will give a comment on another way to describe the photoexcited states of the 1D dimerized Mott insulators.  It is the odd (or even) CT state proposed in Ref.~\onlinecite{okamoto}.  Since the dimerization occurs, a CT state with even parity at $\delta=0$ becomes optically active acquiring a small spectral weight.~\cite{okamoto}  This description is valid for small-$\delta$ systems.  It is a complementary approach to our treatment.  For example, the interdimer CT state adiabatically connects to the lowest photoexcited state of the uniform system, which corresponds to the odd CT state.  
Both the interdimer and intradimer CT states are linear combinations of the odd and even CT states.  As $\delta$ increases, the degree of hybridization changes continuously.  Hence, our discussion from the decoupling limit, which allows us a relatively simple analytical treatment, is helpful to understand the optical excitations of the present system.

Another conclusion of our study is on the origin of the midgap state observed just after the photoexcitation.  Our calculation has clarified that the midgap states correspond to both the intradimer and interdimer CT excitations from $|\psi_{\rm A}\rangle$.  Here we note that the degeneracy between $||E^2\rangle\rangle$ and $|O^3O^1\rangle$ or $|E^3E^1\rangle$ would be lifted for finite $V$ (see Fig.~\ref{fig:opt_for_ktcnq}).  Then the peaks might be split into two parts.
In any case, a dominant spectral weight belongs to the intradimer CT excitation due to the large transfer integral, not to the interdimer CT excitation.
Therefore the observed midgap state in K-TCNQ would be due to the intradimer CT excitation from the photoexcited state $|\psi_{\rm A}\rangle$.

Here we discuss another possibility that the experimentally observed midgap state is not caused by the purely electronic mechanism.
Then the most probable candidate for the midgap state would be a polaron where a photodoped carrier is coupled with phonons via electron-lattice couplings.  In fact, Okamoto et al. have found that the photoinduced phenomena in K-TCNQ are affected by lattice oscillations due to three relevant phonon modes.~\cite{KTCNQ2}  However, it seems inappropriate to explain the midgap state by those phonon modes.  The lowest energy (20cm$^{-1}$) phonon mode has a too-long oscillation period, and the other higher-energy modes decay much faster than the midgap state (which decays in about 3ps).
In this paper, we have provided an idea only on the generation of the midgap state, taking account of
its ultrafast development.  It is a challenging problem to discuss why the midgap state has a relatively long decay time (3ps), although it is beyond the scope of the present study.  A possible scenario is that the midgap state caused by the purely electronic mechanism couples with slowly decaying phonon modes, such as the 20cm$^{-1}$ mode.

\begin{acknowledgments}
The authors are grateful to Prof.~H.~Okamoto for showing his
data prior to publication and for enlightening discussions.
This work was supported by Grants-in-Aid for Creative Scientific Research 
(No.~15GS0216), for Scientific Research on Priority Area ``Molecular 
Conductors'' (No.~15073224), for Scientific Research (C) (No.~15540354), and
NAREGI Nanoscience Project from the Ministry of Education, Culture, Sports, 
Science and Technology, Japan.
Some of numerical calculations were carried out on Altix3700 BX2 at YITP in Kyoto University, and 
on TX-7 at Research Center for Computational Science, Okazaki, Japan.
\end{acknowledgments}

\appendix*
\section{2-site Hubbard model}

The 2-site Hubbard model has the Hamiltonian given by
\begin{equation}
h = -t\sum_{\sigma}( c^\dagger_{1\sigma}c_{0\sigma} +  c^\dagger_{0\sigma}c_{1\sigma} )
 + U ( n_{0\uparrow}n_{0\downarrow}  + n_{1\uparrow}n_{1\downarrow} ),
\label{eq:ham_dimer}
\end{equation}
and the current operator defined by
\begin{equation}
j = it\sum_{\sigma}( c^\dagger_{0\sigma}c_{1\sigma} -  c^\dagger_{1\sigma}c_{0\sigma} ).
\end{equation}
Below we summarize the eigenstates, their energies, and the matrix elements of the current
operator between them, for different numbers of electrons.

\subsection{1-electron states}%

When the dimer contains only one electron, the eigenstates of the Hamiltonian~(\ref{eq:ham_dimer}) for $t>0$ are:
\begin{eqnarray}
|E^1_{\sigma}\rangle &=& \frac{1}{\sqrt{2}} ( c^\dagger_{0\sigma} + c^\dagger_{1\sigma} )|0\rangle  \quad (P=1), \\
|O^1_{\sigma}\rangle &=& \frac{1}{\sqrt{2}} ( c^\dagger_{0\sigma} - c^\dagger_{1\sigma} )|0\rangle  \quad (P=-1).
\end{eqnarray}
In our notation, $E(O)$ indicates that the state has the parity $P=1(-1)$, the superscript shows the number of electrons, and the subscript $\sigma$ is the spin index for the dimer or an electron.
These states obey the eigen equations:
\begin{eqnarray}
h|E^1_{\sigma}\rangle &=& -t|E^1_{\sigma}\rangle, \\
h|O^1_{\sigma}\rangle &=& t|O^1_{\sigma}.\rangle.
\end{eqnarray}
The optical properties are given by
\begin{eqnarray}
j|E^1_{\sigma}\rangle &=& it|O^1_{\sigma}\rangle, \\
j|O^1_{\sigma}\rangle &=& -it|E^1_{\sigma}.\rangle.
\end{eqnarray}

\subsection{3-electron states}%
Next, we consider the case with three electrons.
The eigenstates are 
\begin{eqnarray}
|E^3_{\sigma}\rangle &=& \frac{1}{\sqrt{2}} ( c^\dagger_{0\bar{\sigma}} c^\dagger_{1\sigma}  c^\dagger_{0\sigma} 
- c^\dagger_{1\bar{\sigma}}  c^\dagger_{1\sigma}  c^\dagger_{0\sigma}  ) |0\rangle,   \\
|O^3_{\sigma}\rangle &=& \frac{1}{\sqrt{2}} ( c^\dagger_{0\bar{\sigma}} c^\dagger_{1\sigma}  c^\dagger_{0\sigma} 
+ c^\dagger_{1\bar{\sigma}}  c^\dagger_{1\sigma}  c^\dagger_{0\sigma}  ) |0\rangle ,
\end{eqnarray}
where $\bar{\sigma}=\uparrow(\downarrow)$ for $\sigma=\downarrow(\uparrow)$.
The eigen equations and the optical properties are summarized as 
\begin{eqnarray}
h|E^3_{\sigma}\rangle &=& (U+t)|E^3_{\sigma}\rangle, \\
h|O^3_{\sigma}\rangle &=& (U-t)|O^3_{\sigma}\rangle,
\end{eqnarray}
and
\begin{eqnarray}
j|E^3_{\sigma}\rangle &=& -it|O^3_{\sigma}\rangle, \\
j|O^3_{\sigma}\rangle &=& it|E^3_{\sigma}\rangle.
\end{eqnarray}

\subsection{2-electron spin-singlet states}
Finally, we treat the states containing one electron with up spin and one electron with down spin.
In this subspace, relevant states to photoexcitation are singlet eigenstates summarized as
\begin{eqnarray}
|O^2\rangle &=& \frac{1}{\sqrt{2}} ( c^\dagger_{0 \uparrow} c^\dagger_{0 \downarrow} 
- c^\dagger_{1 \uparrow} c^\dagger_{1 \downarrow}  ) |0\rangle,   \\
|G\rangle &=& \alpha |d\rangle + \beta|c\rangle, \\
|E^2\rangle &=& -\beta |d\rangle + \alpha|c\rangle ,
\end{eqnarray}
where
\begin{eqnarray}
|c\rangle &=& \frac{1}{\sqrt{2}} ( c^\dagger_{0 \uparrow} c^\dagger_{0 \downarrow} 
+ c^\dagger_{1 \uparrow} c^\dagger_{1 \downarrow}  ) |0\rangle, \\
|d\rangle &=& \frac{1}{\sqrt{2}} ( c^\dagger_{0 \uparrow} c^\dagger_{1 \downarrow} 
+ c^\dagger_{1 \uparrow} c^\dagger_{0 \downarrow}  ) |0\rangle, 
\end{eqnarray}
and
\begin{eqnarray}
\alpha &=& 2t/C, \\
\beta &=&  \left[ \sqrt{U^2/4+4t^2 } -   U/2 \right]/C, \\
C&=& \left[U^2/2 + 8t^2 -U\sqrt{U^2/4+4t^2 }            \right]^{1/2}.
\end{eqnarray}
We have $\alpha \to 1$ and $\quad \beta \to 0$ in the $U/t\to \infty$ limit.
Then the ground state $|G\rangle$ approaches $|d\rangle$, which is equivalent to the so-called dimer singlet.
Eigen equations are summarized as follows:
\begin{eqnarray}
h|O^2\rangle &=& U|O^2\rangle, \\
h|G\rangle &=& \epsilon^{-}(t)|G\rangle, \\
h|E^2\rangle &=& \epsilon^{+}(t)|E^2\rangle,
\end{eqnarray}
where
\begin{equation}
\epsilon^{\pm}(t)= U/2 \pm \sqrt{U^2/4+4t^2 }.
\end{equation}
The optical property is given by
\begin{eqnarray}
j|G\rangle &=& 2it\alpha|O^2\rangle, \label{eq:opt_dimer} \\
j|O^2\rangle &=& -2\sqrt{2}it |d\rangle \nonumber \\
&=&  -2\sqrt{2}it( \alpha|G\rangle - \beta|E^2\rangle ).
\end{eqnarray}



\end{document}